\documentclass[aps,prd,twocolumn,preprintnumbers,superscriptaddress,dblfloatfix,nofootinbib]{revtex4-1}

\usepackage[colorlinks,linktocpage,linkcolor=cyan,citecolor=cyan]{hyperref}
\usepackage{multirow}
\usepackage{amsmath,amssymb,mhchem}
\usepackage{graphicx}
\usepackage{dcolumn}
\usepackage{bm}
\usepackage{hyperref}
\usepackage{float}

\newcommand{\liso}{\ensuremath{L_{\rm iso}} }

\begin{document}
\preprint{IPMU22-0065, KEK-QUP-2022-0012, KEK-TH-2473, KEK-Cosmo-0302}
 
\author{Valeri Vardanyan} 
\email{valeri.vardanyan@ipmu.jp}
\affiliation{Kavli Institute for the Physics and Mathematics of the Universe (WPI), UTIAS \\The University of Tokyo, Kashiwa, Chiba 277-8583, Japan}

\author{Volodymyr Takhistov} 
\email{vtakhist@post.kek.jp}
\affiliation{International Center for Quantum-field Measurement Systems for Studies of the Universe and Particles (QUP, WPI),
High Energy Accelerator Research Organization (KEK), Oho 1-1, Tsukuba, Ibaraki 305-0801, Japan}
\affiliation{Theory Center, Institute of Particle and Nuclear Studies (IPNS), High Energy Accelerator Research Organization (KEK), Tsukuba 305-0801, Japan
}
\affiliation{Graduate University for Advanced Studies (SOKENDAI), \\
1-1 Oho, Tsukuba, Ibaraki 305-0801, Japan}
\affiliation{Kavli Institute for the Physics and Mathematics of the Universe (WPI), UTIAS \\The University of Tokyo, Kashiwa, Chiba 277-8583, Japan}

\author{Metin Ata} 
\email{metin.ata@fysik.su.se}
\affiliation{
The Oskar Klein Centre, Department of Physics, Stockholm University, AlbaNova University Centre, SE 106 91 Stockholm, Sweden
}
\affiliation{Kavli Institute for the Physics and Mathematics of the Universe (WPI), UTIAS \\The University of Tokyo, Kashiwa, Chiba 277-8583, Japan}

\author{Kohta Murase} 
\email{murase@psu.edu}
\affiliation{Department of Physics; Department of Astronomy and Astrophysics; Center for Multimessenger Astrophysics, Institute for Gravitation and the Cosmos, The Pennsylvania State University, University Park, Pennsylvania 16802, USA}
\affiliation{School of Natural Sciences, Institute for Advanced Study, Princeton, NJ 08540, USA}
\affiliation{Center for Gravitational Physics and Quantum Information, Yukawa Institute for Theoretical Physics, Kyoto, Kyoto 16802, Japan}

\title{Revisiting tests of Lorentz invariance with gamma-ray bursts:\\ effects of intrinsic lags} 

\begin{abstract} 
Due to their cosmological distances high-energy astrophysical sources allow for unprecedented tests of fundamental physics. Gamma-ray bursts (GRBs) comprise among the most sensitive laboratories for exploring the violation of the central physics principle of Lorentz invariance (LIV), by exploiting spectral time lag of arriving photons. It has been believed that GRB spectral lags are inherently related with their luminosities, and intrinsic source contributions, which remain poorly understood, could significantly impact the LIV results. Using a combined sample of 49 long and short GRBs observed by the \textit{Swift} telescope, we perform a stacked spectral lag search for LIV effects. We set novel limits on LIV, including limits on quadratic effects, and systematically explore for the first time the impacts of the intrinsic GRB lag-luminosity relation. We find that source contributions can strongly impact resulting LIV tests, modifying their limits by up to a factor of few. We discuss constraints coming from GRB 221009A.
\end{abstract}


\maketitle

\section{Introduction}
One of the fundamental pillars of relativity as well as particle physics is the Lorentz invariance symmetry. While extensively confirmed experimentally, speculations exist regarding possible Lorentz invariance violation (LIV), possibly due to certain more fundamental quantum gravity (QG) effects associated with Planck-scale physics~(e.g.,~\citep{Kostelecky:1988zi}). Experimental verification of LIV constitutes an active area of research~\cite{Mattingly:2005re,Kostelecky:2008ts}, with diverse probes ranging from atomic clocks~\citep{Kostelecky:2018fmc} to atmospheric neutrino oscillations~\cite{Super-Kamiokande:2014exs,IceCube:2017qyp}. Due to its central role in modern physics as well as implications for other fundamental themes, such as equivalence principle, it is essential to further extensively explore potential limitations and confront Lorentz invariance with experimental and observational data. 

Distant and energetic astrophysical sources constitute highly sensitive laboratories for exploration of Lorentz invariance. In this regard gamma-ray bursts (GRBs) have been suggested to be particularly useful, and subsequently became a promising frontier; see Ref.~\citep{Amelino-Camelia:1997ieq} for the proposal and e.g.,~\cite{RodriguezMartinez:2006ee,Wei:2021vvn} for reviews on the topic. This can be understood by noticing that while the possible LIV effects appearing at a typical QG energy scale $E_{\rm QG}$ (expected to be of the same order as the Planck-scale  $E_{\rm Pl} = 1.22 \times 10^{19}$~GeV) are highly suppressed at conventional laboratory energies, they could accumulate over cosmological distances due to photon propagation. 
A particularly effective approach for constraining the LIV effects is the investigation of GRB spectral lags, corresponding to different arrival times of photons with different energies due to modified photon dispersion relation in LIV scenarios. Spectral lags are naturally associated with GRB emission pulses~(e.g.,~\cite{Yi:2005ht,Shao:2016mls}).
At cosmological distances, the relatively short spectral lags of energetic GRBs (reaching $\sim 10^{54}$~erg in isotropic-equivalent $\gamma$-ray energy within a few seconds) can be used to probe the LIV-induced photon dispersion~\cite{Amelino-Camelia:1997ieq,Wei:2021vvn}. The characteristic rest-frame photon energies typically range from $\sim100$ keV to $\sim1$ MeV.

One of the central obstacles in such measurements is the robust characterization of the intrinsic astrophysical spectral lags of the sources. Often, simplifying intrinsic lag models are employed, for example assuming that the intrinsic effect is constant and independent of photon energy, and are often generalized to the whole GRB sample in consideration. Distinguishing the intrinsic effect from the propagation effects is also challenging. Lacking a consistent treatment of the intrinsic spectral lags could question the robustness of LIV constraints presented in the literature, and motivates for further studies in this direction.

As an improvement in the intrinsic lag modeling in the context of LIV analyses, in this study, for the first time, we examine systematic effects coming from correlations of GRB lags with their astrophysical properties. Particularly, increasing evidence points towards correlations between intrinsic spectral lags and the GRB luminosities (see e.g.,~\cite{Norris:1999ni,Schaefer:2006pa,Gehrels:2006tk,Ukwatta:2009th,Lu+18}). We reanalyze LIV models using a combined sample of 43 long and 6 short GRBs from \textit{Swift} satellite, and systematically explore how different intrinsic lag prescriptions affect the resulting LIV constrains.

\begin{table*}[htp]
    \centering
    \begin{tabular}{l|c|c|c|c|c|c|c}
    \hline \hline
        GRB ID & $z$ & $\tau$(ms) & $\sigma_{\tau,L}$(ms) & $\sigma_{\tau,R}$(ms) & \liso($10^{51}$~erg/s) & $\sigma_{L_\mathrm{iso}}(10^{51}$~erg/s) & Type \\ \hline
        \hline
        050318 & 1.44 & -13.66 & 184.88 & 218.76 & 4.76 & 0.37 & Long \\ \hline
        050401 & 2.9 & 285.19 & 59.05 & 59.14 & 201.0 & 9.85 & Long \\ \hline
        050525A & 0.61 & 54.72 & 25.42 & 25.59 & 7.23 & 0.18 & Long\\ \hline
        050922C & 2.2 & 162.52 & 74.74 & 79.5 & 184.0 & 28.7 & Long \\ \hline
        060206 & 4.05 & 252.4 & 85.65 & 88.18 & 49.6 & 3.24 & Long  \\ \hline
        060210 & 3.91 & 349.99 & 233.64 & 237.12 & 52.8 & 5.66 & Long \\ \hline
        060306 & 1.55 & 42.56 & 51.17 & 53.73 & 83.0 & 4.9 & Long  \\ \hline
        060814 & 1.92 & -100.01 & 138.04 & 138.73 & 70.9 & 11.7 & Long \\ \hline
        060908 & 1.88 & 230.04 & 169.95 & 175.42 & 12.7 & 1.04 & Long  \\ \hline
        060927 & 5.47 & 14.26 & 111.9 & 111.69 & 108.0 & 7.6 & Long \\ \hline
        061007 & 1.26 & 27.05 & 25.42 & 26.88 & 109.0 & 9.1 & Long  \\ \hline
        061021 & 0.35 & -603.94 & 416.22 & 403.94 & 1.73 & 0.43 & Long  \\ \hline
        061121 & 1.31 & 28.36 & 20.02 & 20.25 & 142.0 & 18.9 & Long \\ \hline
        061222A & 2.09 & 6.07 & 145.67 & 139.01 & 140.0 & 38.0 & Long  \\ \hline
        070521 & 1.35 & 40.2 & 39.51 & 39.07 & 49.3 & 10.9 & Long  \\ \hline
        071020 & 2.15 & 48.47 & 10.7 & 10.24 & 213.0 & 73.0 & Long  \\ \hline
        071117 & 1.33 & 258.54 & 41.21 & 42.58 & 95.3 & 26.2 & Long \\ \hline
        080319B & 0.94 & 30.29 & 21.67 & 19.18 & 102.0 & 9.4 & Long  \\ \hline
        080319C & 1.95 & 217.82 & 168.48 & 171.2 & 96.1 & 21.2 & Long  \\ \hline
        080413B & 1.1 & 96.0 & 61.91 & 59.56 & 14.9 & 0.62 & Long \\ \hline
        080603B & 2.69 & -43.59 & 67.38 & 63.01 & 116.0 & 30.0 & Long  \\ \hline
        080605 & 1.64 & 53.65 & 36.46 & 37.38 & 308.0 & 64.0 & Long  \\ \hline
        080607 & 3.04 & 90.99 & 91.44 & 101.78 & 2259.0 & 453.0 & Long  \\ \hline
        080721 & 2.59 & -158.16 & 162.73 & 149.69 & 1038.0 & 172.0 & Long \\ \hline
        080804 & 2.2 & -347.4 & 618.25 & 623.99 & 27.0 & 3.3 & Long  \\ \hline
        080916A & 0.69 & 599.82 & 288.57 & 290.73 & 1.08 & 0.06 & Long  \\ \hline
        081121 & 2.51 & -10.41 & 245.62 & 266.41 & 195.0 & 31.0 & Long  \\ \hline
        081203A & 2.1 & -39.23 & 198.37 & 175.09 & 28.2 & 1.9 & Long  \\ \hline
        081221 & 2.26 & 99.44 & 77.55 & 80.56 & 100.0 & 2.0 & Long \\ \hline
        081222 & 2.77 & 129.02 & 81.04 & 86.36 & 94.9 & 3.1 & Long  \\ \hline
        090102 & 1.55 & 522.53 & 278.44 & 304.17 & 45.7 & 1.4 & Long  \\ \hline
        090424 & 0.54 & 18.62 & 47.22 & 50.44 & 11.2 & 0.17 & Long  \\ \hline
        090715B & 3.0 & 70.66 & 304.24 & 385.39 & 82.6 & 22.9 & Long \\ \hline
        090812 & 2.45 & 168.71 & 338.84 & 343.29 & 96.2 & 9.7 & Long  \\ \hline
        090926B & 1.24 & 1031.73 & 861.13 & 887.57 & 4.28 & 0.25 & Long  \\ \hline
        091018 & 0.97 & 163.65 & 147.37 & 149.05 & 4.73 & 1.04 & Long \\ \hline
        091020 & 1.71 & -78.58 & 282.06 & 290.03 & 32.7 & 4.6 & Long \\ \hline
        091127 & 0.49 & 157.64 & 194.65 & 192.49 & 9.08 & 0.22 & Long  \\ \hline
        091208B & 1.06 & 84.2 & 31.61 & 31.6 & 17.4 & 0.7 & Long  \\ \hline
        100621A & 0.54 & 924.74 & 727.39 & 677.68 & 3.17 & 0.24 & Long  \\ \hline
        100728B & 2.11 & -115.0 & 456.44 & 406.26 & 18.7 & 1.2 & Long  \\ \hline
        110205A & 2.22 & -125.63 & 136.21 & 144.66 & 25.1 & 3.4 & Long  \\ \hline
        110503A & 1.61 & 46.77 & 82.15 & 85.65 & 181.0 & 18.0 & Long  \\ \hline
        051221A & 0.55 & -1.85 & 2.32 & 2.47 & 58.4 & 8.9 & Short  \\ \hline
        070714B & 0.92 & 5.58 & 35.01 & 31.56 & 13.0 & 1.4 & Short \\ \hline
        090510 & 0.9 & -7.99 & 8.4 & 8.63 & 178.0 & 11.7 & Short  \\ \hline
        101219A & 0.72 & -0.02 & 21.77 & 22.42 & 65.0 & 18.6 & Short  \\ \hline
        111117A & 1.3 & 3.24 & 10.7 & 10.1 & 40.4 & 12.8 & Short  \\ \hline
        130603B & 0.36 & -3.44 & 5.58 & 7.27 & 43.5 & 8.7 & Short \\ \hline \hline
    \end{tabular}
    \caption{\label{tab:data} The combined data set used in this work. The most left column indicates the GRB's unique identifier,   followed by the GRB's redshift estimate. The next column shows the observed spectral lag between two energy bands, where $\sigma_{\tau,L/R}$ are the left and right-sided measurement errors. \liso and $\sigma_{L_\mathrm{iso}}$ indicate the GRBs' bolometric isotropic-equivalent luminosity and its $1~\sigma$ uncertainty. The most left column indicates if the GRB was a long or short pulse.}
\end{table*}

\begin{figure*}[tb]\label{fig:data}
\begin{center}
\includegraphics[clip,width=\columnwidth]{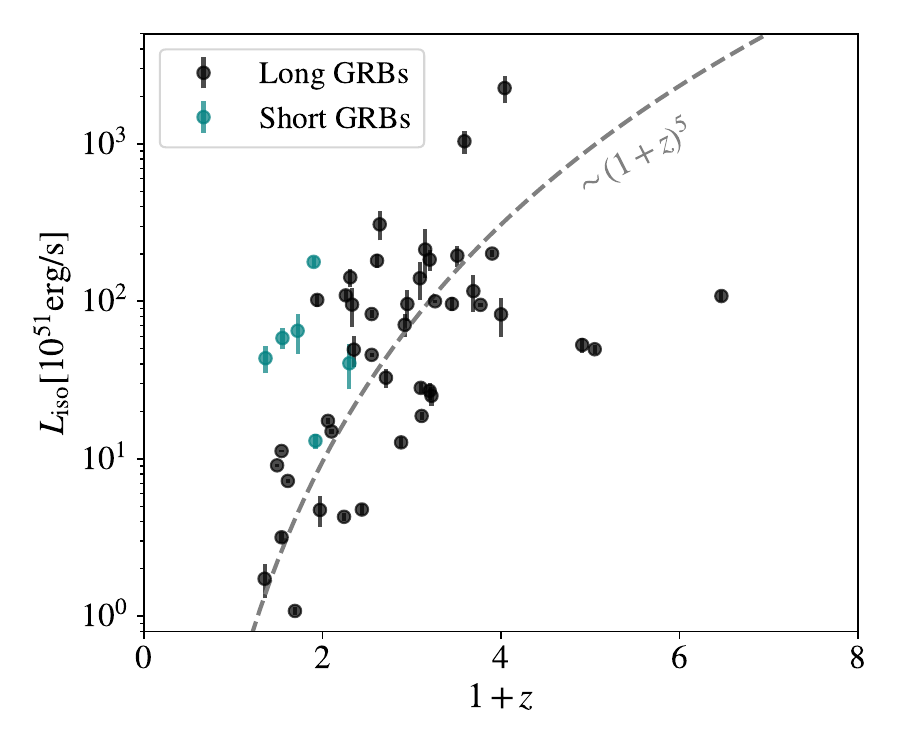}
\includegraphics[clip,width=\columnwidth]{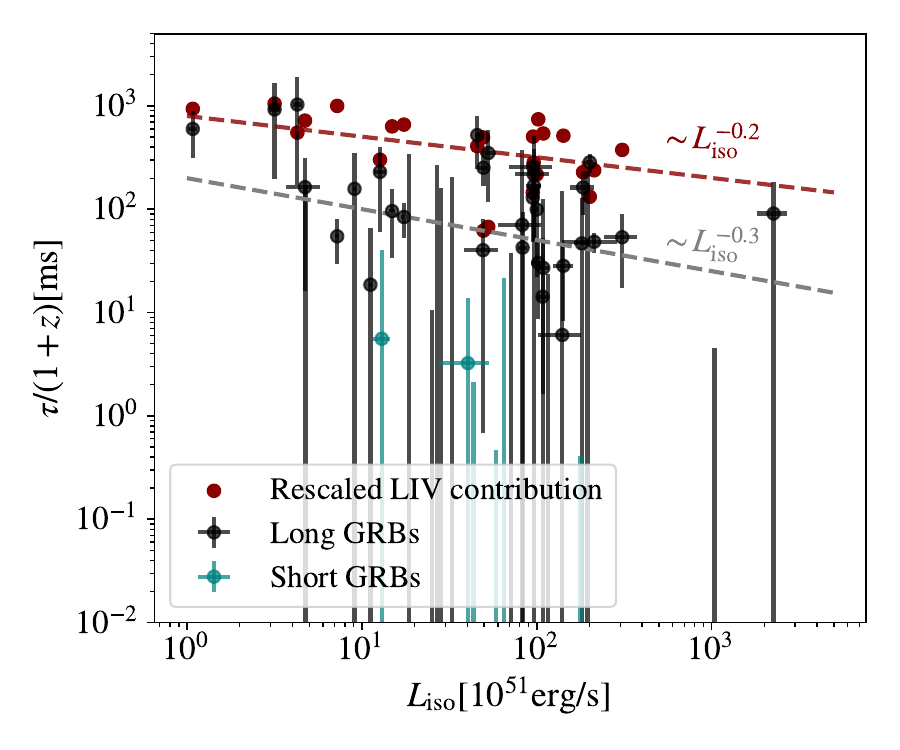}
    \caption{Relationship between spectral time-lag and luminosity of 43 long and 6 short GRBs detected by \textit{Swift}. \textbf{Left:} Redshift evolution of GRB luminosities. The dashed gray line roughly depicts the median of the data points, and shows that luminosities are correlated with redshift. \textbf{Right:} Lag-luminosity correlations similar to the results presented in Ref.~\cite{Bernardini2015}. The dashed gray line roughly depicts the median of positive-lag GRBs. The red points depict the arbitrarily rescaled contributions from LIV physics at the redshift of the particular positive-lag GRB. The red, dashed line roughly corresponds to the median of the red points. See text for more details.}
\end{center}
\end{figure*}

\section{Gamma-ray Burst Spectral lags}
\label{sec:grblag}
GRBs exhibit well-known bimodality in distribution of their pulse duration, and can be broadly characterized as ``long'' with duration of $\gtrsim 2$~s and ``short'' with duration of $\lesssim 2$~s~\cite{Kouveliotou:1993yx}. This behaviour could be attributed to their possibly distinct origins; the long GRBs could be associated with collapsar massive stars~\cite{Woosley:1993wj}, while the short ones with binary mergers including neutron stars~(see e.g.,~\cite{Paczynski:1986px,Narayan:1992iy}). 
The latter association is supported by the multi-messenger observations of a binary neutron star (events GRB 170817A and GW170817) \cite{LIGOScientific:2017vwq,LIGOScientific:2017ync,LIGOScientific:2017zic}. Intriguingly, long GRBs often exhibit positive spectral lags, i.e., the arrival times of emitted high-energy photons precedes those of low-energy photons, while short GRBs are generally consistent with a negligible spectral lag~(e.g.,~\cite{Norris:1999ni,Norris:2006rw,Ukwatta:2009th}). It should be noted, however, that exceptions to this approximate classification have been pointed out (e.g.,~\cite{Gehrels:2006tk}). Following the original analysis of Ref.~\citep{Norris:1999ni} that employed six BATSE GRBs, variety of subsequent studies have provided further evidence that spectral lags are anti-correlated with the isotropic-equivalent peak luminosity~(e.g.,~\citep{Schaefer:2006pa,Gehrels:2006tk,Ukwatta:2009th}).

GRB pulses are primarily observed with positive spectral lags. However, GRBs with zero or even negative spectral lags have also been detected~(e.g.,~\cite{Ukwatta:2009th,Ukwatta:2011ui,Roychoudhury:2014,Lu+18}). Variety of mechanisms have been suggested to explain the positive spectral lags, including curvature effects \citep{Nakamura:2001kd,Shen+05,Lu+06}, the intrinsic cooling of radiating electrons~\citep{Schaefer:2004}. It is not straightforward to explain the observed lags even with detailed studies taking into account both hydrodynamical and radiative effects~\citep{DM03,BD14}, and accelerating outflows with decreasing magnetic fields have also been suggested~\citep{UZ16,UZ18,Du:2019ees}.    
While negative spectral lags are more challenging to describe theoretically, a possible explanation includes the up-scattering of soft radiation via inverse Compton emission~\citep{Ryde:2004wf}. 
Many models for spectral lags involve spectral evolution, and soft-to-hard spectral transitions of a source could be responsible for the negative lags~\citep{CHAKRABARTI:2018}.

\section{PHOTON VACUUM DISPERSION AND LORENTZ INVARIANCE}
Assuming the Lorentz invariance, special relativity predicts the dispersion relation between photon energy $E$ and momentum $p$ to be $E^2 = p^2 c^2$. LIV would result in energy-dependent speed of light in vacuum, leading to a modified dispersion relation. In this work, instead of focusing on a first-principle derivation of the dispersion relation we consider a phenomenological parametrization of the LIV effects. Particularly, we consider the following modified photon dispersion relation
\begin{equation}
    E^2 \simeq p^2 c^2 \Big[1 - \sum_{n = 1}^{\infty} s_{\rm \pm}\Big(\dfrac{E}{E_{\rm QG,n}}\Big)^n\Big]~,
\end{equation}
where $E_{\rm QG,n}$ is the characteristic energy scale where QG effects become relevant. For $E \ll E_{\rm QG}$ lower order terms dominate, and linear ($n = 1$) or quadratic ($n = 2$) LIV effects can be considered (e.g.,~\cite{Horava:2008ih,Vacaru:2012nvq,Sefiedgar:2010we}).
The resulting modified photon propagation speed (group velocity) is given by
\begin{equation} \label{eq:ve}
    v(E) = \dfrac{\partial E}{\partial p} \simeq c \Big[1 - s_{\rm \pm}\dfrac{n+1
    }{2}\Big(\dfrac{E}{E_{\rm QG,n}}\Big)^n\Big]~.
\end{equation}
The coefficient $s_{\rm \pm}$ represents the sign of LIV effects, corresponding to
the ``subluminal'' $(s_{\rm \pm} = +1)$ scenario with decreasing photon speed or the ``superluminal'' $(s_{\rm \pm} = -1)$ scenario with increasing photon speed as the photon energy increases.

Because of the energy dependence of $v(E)$, simultaneously emitted high- and low-energy photons with observer-frame energies $E_h$ and $E_l$ ($E_h > E_l$) reach the observer at different times.
The delay in photon arrival time due to LIV is given by~\cite{Jacob:2008bw}
\begin{align}\label{eq:LIV_lag}
\Delta \tau_{\rm LIV} =&~ \tau_\mathrm{h} - \tau_\mathrm{l} \\
=&~ s_{\rm \pm} \dfrac{1+n}{2 H_0} \dfrac{E_h^n - E_l^n}{E_{\rm QG,n}^n} \int_0^z \dfrac{(1 + z^{\prime})^n dz^{\prime}}{\sqrt{\Omega_m ( 1 + z^{\prime})^3 + \Omega_{\Lambda}}}~, \notag
\end{align}
where $\tau_\mathrm{h}$ and $\tau_\mathrm{l}$ are arrival times of the high- and low-energy photons, respectively, $z$ is the redshift of the source, and cosmological parameters of the standard $\Lambda$CDM model are fixed as $H_0 = 70$~km s$^{-1}$ Mpc$^{-1}$, $\Omega_{\rm m} = 0.27$ and $\Omega_{\Lambda} = 0.73$. We note that $s_{\rm \pm} = +1$ leads to LIV-induced
negative time lags. 

As a result, in the presence of LIV effects, the total delay in photon arrival time between distinct energy bands is given by
\begin{equation}
\Delta \tau = \Delta \tau_{\rm int} + \Delta \tau_{\rm LIV}~.
\end{equation}
Furthermore, additional contribution $\Delta \tau_{\rm DM}$ could appear due to arrival-time differences caused by from the dispersion by the line-of-sight free electron distribution. Additionally, other new-physics effects such as time delay $\Delta \tau_{\rm pm}$ due to non-zero photon mass and a delay $\Delta \tau_{\rm ep}$ due to violation of Einstein's equivalence principle could also alter the interpretation of $\Delta \tau$. However, recent analyses suggest that these contributions do not significantly impact GRB photons~(see Refs.~\cite{Wei:2015hwd,Gao:2015lca} for earlier analyses, and Refs.~\cite{Bartlett:2021yyp,Hashimoto:2021swe,Minazzoli:2022lzs} for cosmologically consistent treatments). In this paper, we assume that $\Delta \tau_{\rm int}$ and $\Delta \tau_{\rm LIV}$ are the dominant contributions to $\Delta \tau$.

\begin{figure*}[t]
    \centering
\includegraphics[clip,width=\columnwidth]{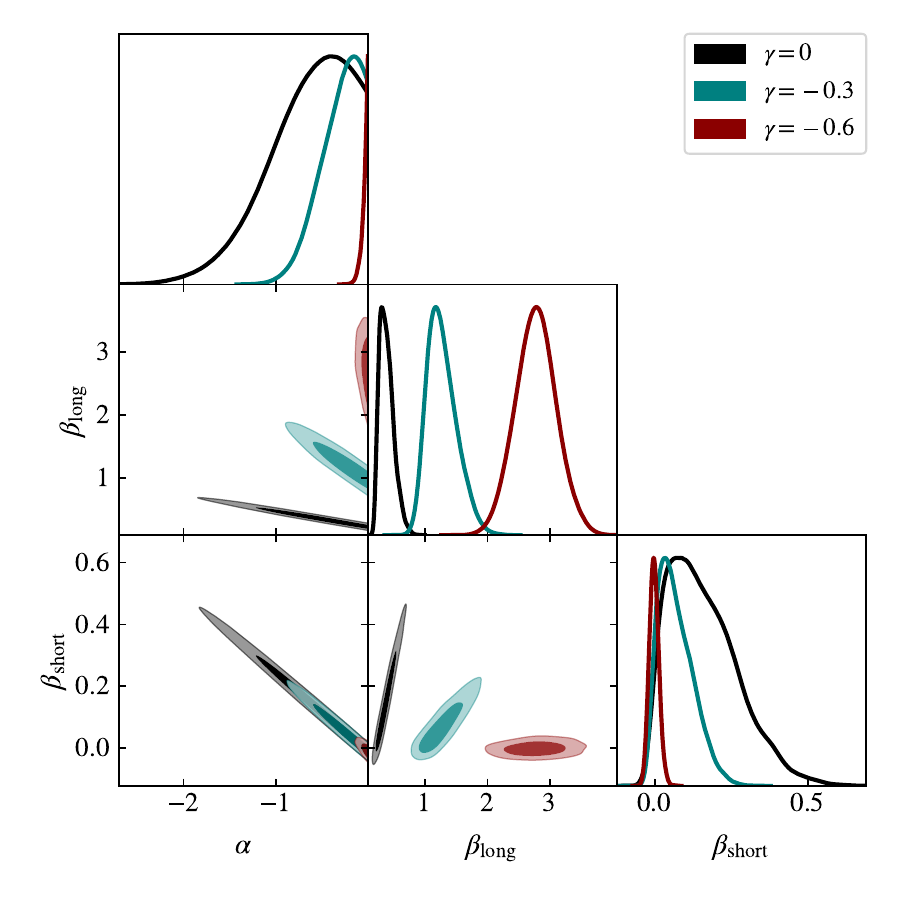} 
\includegraphics[clip,width=\columnwidth]{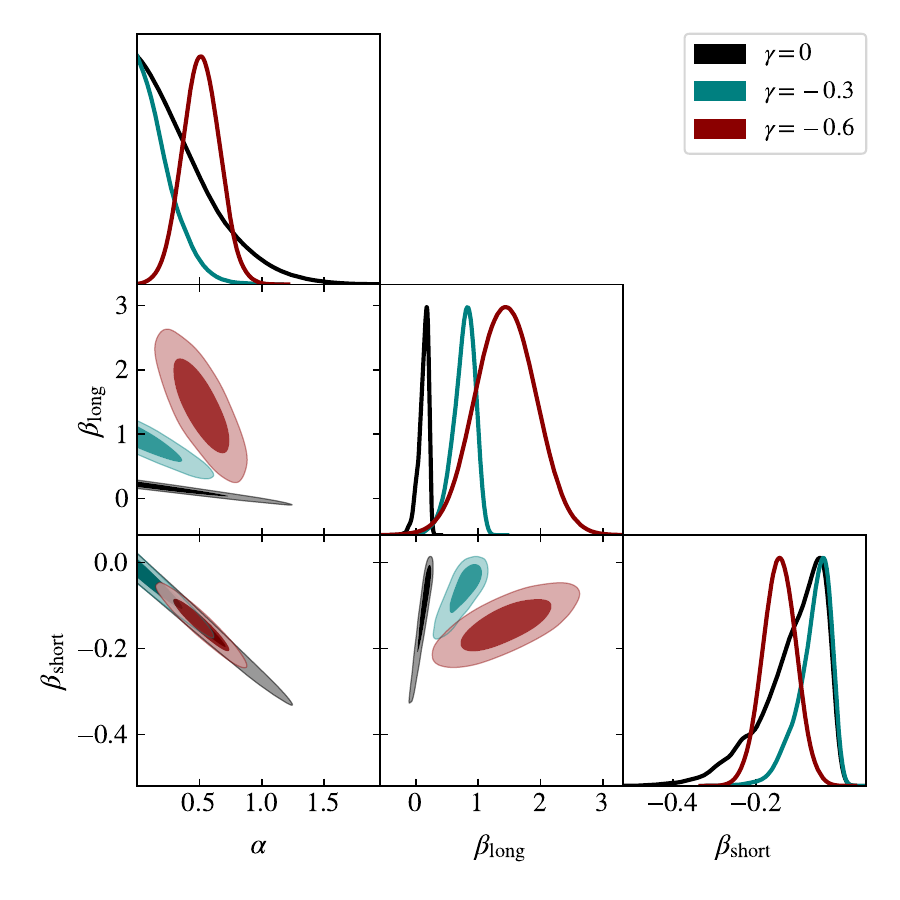}
    \caption{Posterior distributions of intrinsic model parameters $\beta_\mathrm{long}$ and $\beta_\mathrm{short}$, and the LIV parameter $\alpha$ defined in Eq.~(\ref{eq:alpha}).  \textbf{Left:} The case of $s_\pm = -1$. The $95 \%$ limits on $\alpha$ are $>-1.49, >-0.72, >-0.11$ for $\gamma = 0, -0.3$ and $-0.6$, respectively.  \textbf{Right:} The case of $s_\pm = +1$. The $95 \%$ limits on $\alpha$ are $<0.98, <0.49, 0.51^{+0.30}_{-0.29}$ for $\gamma = 0, -0.3$ and $-0.6$, respectively.}
    \label{fig:limits}
\end{figure*}

\section{Data Set}
In our analysis we employ a data set of 50 long and 6 short bright GRBs detected by \textit{Swift}/BAT instrument and compiled in \citet{Bernardini2015}, which also provides the redshifts and spectral lags of the sources. The sample covers GRBs across broad redshift range from $0.35$ to $5.47$. Furthermore, we utilize the bolometric isotropic-equivalent luminosity \liso measurements provided in \citet{Nava2012} (for long GRBs) and \citet{DAvanzo:2014urr} (for short GRBs). In order to make our paper self-contained, we provide the combined data in Table~\ref{tab:data}. Out of the 50 long GRBs in the original catalogue we only use the 43 sources described in Table~\ref{tab:data}, since only those sources have unique redshifts  and luminosity measurements. Therefore, our final sample includes 49 sources in total. Table~\ref{tab:data} specifies the GRB ID, redshift $z$, spectral lag $\tau$ between two energy bands, left-sided and right-sided measurement errors on the spectral lag $\sigma_{\tau, L/R}$, GRB isotropic-equivalent luminosity $L_{\rm iso}$ and its uncertainty $\sigma_{L_{\rm iso}}$ as well as a flag specifying whether the GRB is classified as long or short.

The source rest-frame energy $E^{\prime}$ is related to observer-frame energy $E$ through $E^{\prime} = E ( 1 + z)$. Using the appropriate observer-frame energy bands based on the redshift of each GRB, Ref.~\cite{Bernardini2015} extracted time lags between characteristic rest-frame energy bands of $100–150$
and $200–250$ keV. In our analysis the energy gap between mid-points of rest-frame bands is therefore fixed for each source to be $\Delta E^{\prime} = 100$~keV, while the observer-frame energy gaps vary depending on the source redshift. Constant observer-frame energy gaps have been employed in e.g.,~\cite{Ukwatta:2011ui}.

\section{Analysis and Results}
Correct modeling of intrinsic GRB time-lags is important for reliably inferring constraints on LIV physics. In order to account for unknown intrinsic lag $\Delta t_{\rm int}$, most previous studies have assumed a constant rest-frame value $b$, universal for all GRBs; $\Delta t_{\rm int} = b(1 + z)$ (see e.g.,~\cite{Ellis:1999sd,Ellis:2005sjy,Pan:2015cqa}). An improved and more general analysis has been carried out in \cite{Bartlett:2021olb}, where the intrinsic lags are described with a Gaussian mixture model. Alternative \textit{ad hoc} models included assuming a power-law energy dependence for the time-lag between the lowest energy band and any other higher energy bands~\cite{Wei:2016exb}. Treatment of source effects based on just random number statistics have also been proposed~\cite{Ellis:2018lca}.

In this paper, we take into account the correlations between the time-lag and peak GRB luminosities. Such a correlation has been found empirically for long GRBs (e.g.,~\cite{Norris:1999ni,Norris:2006rw,Ukwatta:2009th}), and is predicted by theoretical models (e.g.,~\cite{Nakamura:2001kd, DM03, Schaefer:2004, UZ16}). 
GRB luminosities are primary physical observables, and they can be indispensable for separating source-specific, intrinsic time-lags and lags originating due to photon propagation. This provides a physically motivated model of the intrinsic time-lag, therefore allowing us to derive more robust constraints on the LIV parameters than what has been previously obtained. Here, for the first time, we implement and explore lag-luminosity relation for intrinsic lag in the context of LIV constraints. As discussed in Section~\ref{sec:grblag}, lag-luminosity correlations are primarily expected for long GRB populations.  

For our LIV analysis, for the long GRBs we model the intrinsic rest-frame lag as
\begin{equation}
    \tau_\mathrm{RF}^{\mathrm{int}, i} = \frac{\tau_\mathrm{obs}^{\mathrm{int}, i}}{1 + z} = \beta_\mathrm{long}\left(\frac{L_\mathrm{iso}^{i}}{L_\ast}\right)^\gamma,
\end{equation}
where $i$ labels the long GRBs in our catalogue, $\beta_{\rm long}$ and $\gamma$ are free parameters, and $L_\ast$ is an arbitrary normalization scale. In our analysis, we also include the short GRBs. For this population, however, lacking empirical and theoretical motivations for the intrinsic lag-luminosity correlations, we have assumed a uniform constant rest-frame time-lag: $\tau_\mathrm{RF}^{\mathrm{int}, j} = \beta_\mathrm{short}$, where $j$ labels the individual short GRBs in our catalogue and $\beta_\mathrm{short}$ is a free constant parameter.

For the LIV contribution, we find that the optimal quantity to constrain is the combination
\begin{align}
    \alpha \equiv s_{\rm \pm} \dfrac{1}{H_0} \dfrac{E_h^{\prime n} - E_l^{\prime n}}{E_{\rm QG,n}^n}~.
    \label{eq:alpha}
\end{align}
We note that for a fixed $\Delta E^{\prime n}$, $\Delta E^n = E_h^n -E_l^n$ varies depending on the source redshift.

With our primary interest being the relationship between the LIV time-lag contributions and the intrinsic astrophysically-induced time-lags, it is important to note that long GRB luminosities  vary with redshift as it is demonstrated in the left panel of Fig.~\ref{fig:data}. On the other hand, since in our model the intrinsic time-lags are tightly connected to luminosities, the rest-frame intrinsic time-lags also vary with redshift. This redshift dependence partly degenerates with the redshift dependence of the LIV prediction. Indeed, because of its redshift dependence, the LIV contributions are indirectly correlated with GRB luminosities. As we demonstrate in the right panel of Fig.~\ref{fig:data}, this correlation cannot account for the entire observed lag-luminosity correlation, therefore suggesting the usefulness of using the luminosity information in our analysis. This conclusion is supported by noting that the red-dashed line, roughly corresponding to the median of contributions from LIV physics at the redshifts of the positive-lag GRBs, differs noticeably from the median of the observed positive-lag data points (roughly depicted by the gray-dashed line in the same figure). Here, we conservatively assume that possible additional sources of uncertainties do not significantly affect these results.  

\begin{table*}[ht]
    \centering
    \begin{tabular}{l|c|c}
    \hline \hline
        LIV model & Lag-luminosity $\gamma$  & 95 \% C.L. limit on $E_{\rm QG}[\textrm{GeV}]$ \\ 
        \hline
        \multirow{3}{7em}{$n = 1, s_{\pm} = +1$}
          & $0.0$ & $> 4.47\times 10^{14}$ \\  
          & $-0.3$ & $> 8.95\times 10^{14}$ \\  
          & $-0.6$ & $E_{\rm QG}/10^{14}=8.6^{+2.4}_{-3.2}$\\ 
        \hline
         \multirow{3}{7em}{$n = 1, s_{\pm} = -1$} & $0.0$ & $> 2.95 \times 10^{14}$ \\
          & $-0.3$ & $> 6.10 \times 10^{14}$ \\ 
          & $-0.6$ & $> 4.01\times 10^{15}$ \\ 
\hline
\hline
         \multirow{3}{7em}{$n = 2, s_{\pm} = +1$} & $0.0$ & $> 4.14\times 10^{5}$ \\  
          & $-0.3$ & $> 5.75\times 10^{5}$ \\  
          & $-0.6$ & $E_{\rm QG}/10^5=5.6^{+3.5}_{-1.2}$ \\ 
        \hline
         \multirow{3}{7em}{$n = 2, s_{\pm} = -1$} & $0.0$ & $> 3.46\times 10^{5}$ \\ 
          & $-0.3$ & $> 4.03\times 10^{5}$ \\ 
          & $-0.6$ & $> 1.13\times 10^{6}$ \\
        \hline
        \hline  

    \end{tabular}
    \caption{\label{tab:results} Summary of $95 \%$ confidence intervals for the LIV parameter $E_{\rm QG}$ for the considered models. Note that for most of the models, data can only provide lower bounds. In some cases, however, both lower and upper confidence intervals can be identified. This emphasizes the importance of detailed modeling of intrinsic lags.}
\end{table*}

Our objective is to estimate the effect of the $\gamma$ parameter on the inference of the quantum gravity scale $E_\mathrm{QG}$. In practice, we explore the $\alpha$ parameter alongside the $\beta_\mathrm{long}$ and $\beta_\mathrm{short}$ parameters for a selection of fixed values of $\gamma$. We assume all the time-lag measurement errors to be independent, and construct a Gaussian likelihood of the form
\begin{align}
    \mathcal{L} = \prod_i e^{-(\tau^i_{\rm theor} - \tau^i_{\rm obs})^2/2{\sigma^i_\tau}^2},
\end{align}
where $i$ is the data index, subscript ``$\rm theor$'' denotes the theoretical prediction, including both the intrinsic and LIV contributions, subscript ``$\rm obs$'' denotes the data, $\sigma^i_\tau$ is approximated by the average of $\sigma_{\tau, \rm{L}}$ and $\sigma_{\tau, \rm{R}}$ in Table~\ref{tab:data}. We have checked that taking into account the luminosity uncertainties does not noticeably alter our inference.   

The posterior distribution is sampled using \texttt{emcee} \citep{ForemanMackey:2012ig}, and the statistical summaries are calculated using \texttt{GetDist} \citep{getdist}. Since $s_\pm$ is a discreet parameter, we perform inference for $s_\pm \pm 1$ cases separately. Our results are summarized in Figure~\ref{fig:limits} for the case of $n = 1$ and $s_\pm = -1$ (left panel), and $s_\pm = +1$ (right panel). From Eq.~(\ref{eq:alpha}) it should be noted that for $s_\pm = -1$, the parameter $\alpha$ can only take negative values, hence, its prior should be bounded from above by zero. Similarly, for $s_\pm = +1$, the parameter $\alpha$ can only take positive values. The results for $n = 2$ look very similar, and we choose not to show the full posteriors for this case. This similarity is expected since $n$ only mildly changes the redshift integrand in Eq.~(\ref{eq:LIV_lag}).

Previously, the data set of Ref.~\cite{Bernardini2015} was already employed for LIV analysis by Ref.~\cite{Wei:2017qfz}, without taking into account lag-luminosity relation and assuming constant intrinsic lag. By considering $\gamma = 0$  and $s_{\pm} = -1$ case in our analysis, our resulting limits are compatible with results presented in Ref.~\cite{Wei:2017qfz}.
Figure~\ref{fig:limits} demonstrates the impact of intrinsic modeling on the inferred constraints. These results particularly emphasize that the choice of the astrophysical model can change the conclusions about the LIV physics, ranging from stringently constraining it, all the way to detecting hints of its presence. While not our main objective in this work, clearly, when marginalizing over $\gamma$, one would obtain the most conservative constraints, corresponding to broader $\alpha$ distributions compatible with $0$.

Precise determination of photon spectra time-dependence could further help distinguish source and LIV effects~\cite{Amelino-Camelia:1997ieq}. Future surveys and observations, such as by the proposed THESEUS satellite~\cite{theseus:2021}, will be able to probe LIV effects near the Planck scale.

\section{Conclusions}
\label{sec:conclusions}
Energetic and distant astrophysical sources make it possible to test fundamental physics at energies far exceeding the capabilities of terrestrial laboratories. GRBs constitute excellent sources to explore the possible violations of fundamental principle of Lorentz invariance. However, the inferred LIV limits could be significantly affected by poorly understood intrinsic source effects that are often modeled inconsistently. In this work, using long and short GRB samples from \textit{Swift} telescope, we explored the impacts of the source luminosities and their correlations with intrinsic spectral lags on LIV measurements. We have demonstrated the effect of intrinsic lag modelling on the derived constraints, and have shown that incorrect intrinsic models can lead to qualitatively incorrect or aggressive conclusions regarding the LIV physics. Additionally, we have derived the first limits on quadratic LIV effects using the \textit{Swift} data set.

The LIV limits from MeV $\gamma$-rays, on which we focused in this work, are complementary to those from higher-energy $\gamma$-rays. GRBs are known to be not only MeV emitters but also GeV-TeV emitters that can also be used as a probe of higher scales of LIV effects, and even ultrahigh-energy $\gamma$-rays have been proposed as a test of the LIV effects~\citep{Murase:2009ah}. 
In particular, one of the short GRBs, GRB 090510, detected by \textit{Fermi}-Large Area Telescope (LAT) placed a stringent LIV limit, $E_{\rm QG} > 9.3 \times 10^{19}~$GeV~\cite{abdo:2009,Vasileiou:2013vra}. 
The recently detected unusually bright burst event GRB 221009A, registered on October 9, 2022 at redshift $z = 0.151$~\cite{Postigo:2022}. The event has been detected by \textit{Swift}-BAT as an unknown-type transient \cite{Dichiara:GCN} with \textit{Fermi}-GBM observing it an hour before the BAT trigger \cite{Veres:GCN}. 
The detection of a $\sim100$~GeV photon may provide a weaker constraint via the LIV-induced time delay~\citep{Zhu:2022usw}. 
The detection of very high-energy photons up to $\sim18$~TeV by LHAASO~\cite{Huang:GCN} has also been of interest, because LIV effects could significantly modify photon dispersion relations and propagation opacity ~\cite{Galanti:2022pbg,Li:2022wxc,Finke:2022swf, Zheng:2022ooe,Baktash:2022gnf}. The scale of LIV effects, $E_{\rm QG} \lesssim 1 \times 10^{20}~$GeV (assuming $n = 1$) is allowed by the limits from our \textit{Swift} data analysis (see Tab.~\ref{tab:results}).
On the other hand, such effects are already under mild pressure from limits of $E_{\rm QG} \gtrsim 1 \times 10^{20}~$GeV obtained by stacking analysis of GRB spectra using data sets from HEGRA, HESS, VERITAS, TACTIC, ARGO-YBJ, Whipple observations~\cite{Lang:2018yog}. 
However, we caution that the LIV limits relying on the absorption effect requires details of not only the very high-energy photon data but also the GRB modeling~\cite{Zhao:2022wjg,Zhang:2022lff}. In addition, GeV-TeV $\gamma$-rays do not have to share the origin with MeV $\gamma$-rays. Rather, they may be produced by afterglows, in which the constraints are subject to different systematics. Our limits using the MeV $\gamma$-ray data are important as conservative and complementary constraints.

~\newline
\acknowledgments

We thank Kunihito Ioka, Kazunori Kohri, Ruben Salvaterra for valuable discussions. This work was supported, in part, by the World Premier International Research Center Initiative (WPI), MEXT, Japan. The work was partly supported by JSPS KAKENHI grant No.~20K22348 (V.V.), No. 23K13109 (V.T.), JSPS KAKENHI Grant Number JP21K13911 (M.A.), NSF Grants No.~AST-1908689, No.~AST-2108466 and No.~AST-2108467, and JSPS KAKENHI grants No.~20H01901 and No.~20H05852 (K.M.)
 
\bibliography{bibliography}

\end{document}